\documentclass[pra,aps,showpacs,groupedaddress,superscriptaddress,twocolumn,toc=flat,biblatex,footinbib]{revtex4-1}
\usepackage{natbib}
\usepackage[table]{xcolor}
\usepackage[colorlinks=false]{hyperref}
\usepackage[utf8]{inputenc}
\usepackage{color}
\usepackage{bbm} 
\usepackage[english]{babel}
\usepackage{multirow}
\usepackage{amsfonts,amsmath,amssymb,stmaryrd}

\usepackage{graphicx}
\usepackage{subfigure}  

\usepackage{epsfig}
\usepackage{mathrsfs}
\usepackage{verbatim}
\usepackage{centernot}
\usepackage{ulem}
\usepackage{longtable}
\usepackage{nicefrac}
\usepackage[framemethod=tikz]{mdframed}
\usepackage{siunitx}
\usepackage[nolist]{acronym}
 
\newcommand{\mycomment}[1]{} 

\usepackage{array}

\usepackage{cancel,ifthen}
\newcommand{\cmnt}[2][NoInPuT]{\ifthenelse{\equal{#1}{NoInPuT}}{}{{\color{red}\sout{#1}}} {\color{blue} #2}}

\usepackage{bm}	
\renewcommand{\vec}[1]{\bm{#1}}

\LTcapwidth=\textwidth

\begin{document}

\normalem	

\title{Defect-enhanced diffusion of magnetic skyrmions}

\author{Philipp Rieger}
\email[]{philipp.rieger@uni-konstanz.de}

\affiliation{Department of Physics, University of Konstanz, DE-78457 Konstanz, Germany}

\author{Markus Weißenhofer}
\affiliation{Department of Physics, University of Konstanz, DE-78457 Konstanz, Germany}
\affiliation{Department of Physics, Free University Berlin, Arnimallee 14, 14195 Berlin, Germany}
\affiliation{Department of Physics and Astronomy, Uppsala University, P. O. Box 516, S-751 20 Uppsala, Sweden}

\author{Ulrich Nowak}
\affiliation{Department of Physics, University of Konstanz, DE-78457 Konstanz, Germany}


\date{\today}

\begin{abstract}
Defects, i.e. inhomogeneities of the underlying lattice, are ubiquitous in magnetic materials and can have a crucial impact on their applicability in spintronic devices. For magnetic skyrmions, localized and topologically non-trivial spin textures, they give rise to a spatially inhomogeneous energy landscape and can lead to pinning, resulting in an exponentially increased dwell time at certain positions and typically a strongly reduced mobility. Using atomistic spin dynamics simulations, we reveal that under certain conditions defects can instead enhance thermal diffusion of ferromagnetic skyrmions. By comparing with results for the diffusion of antiferromagnetic skyrmions and using a quasi-particle description based on the Thiele equation, we demonstrate that this surprising finding can be traced back to the partial lifting of the impact of the topologigal gyrocoupling, which governs the dynamics of ferromagnetic skyrmions in the absence of defects.
\end{abstract}

\maketitle


\section{Introduction}

    Magnetic skyrmions are topologically protected spin configurations where the directions of the magnetic moments span the whole unit sphere, forming a nanoscale particle-like object with finite topological charge $Q= \frac{1}{4\pi}\int \vec{n}\cdot (\partial_x \vec{n} \times \partial_y \vec{n})\ \mathrm{d}^2r$, with $\vec{n}$ being the magnetic order parameter \cite{Bogdanov_1994,Muehlbauer2009,Nagaosa2013}. 
    Due to their small size, robustness, controllable creation and annihilation \cite{doi:10.1126/science.1240573}, and high mobility \cite{Jiang2017, Litzius2017}, skyrmions are suitable candidates for information carriers in future nanoscale magnetic logic and memory devices \cite{Fert2013,Zhang2015, 10.1093/nsr/nwy109}. The observation of thermally-induced Brownian motion \cite{reshuffler, doi:10.1063/1.5070101, PhysRevLett.125.027206} has also attracted considerable attention because of its possible application for probabilistic computing \cite{reshuffler,PhysRevApplied.9.064018, doi:10.1063/5.0011105}. 
    
    However, designing effective skyrmion devices requires an thorough understanding of the interactions between skyrmions and defects, as defects are almost unavoidable in magnetic materials and can significantly affect the motion of skyrmions. Previous works have demonstrated that defects, such as vacancies \cite{PhysRevB.91.054410}, enhanced  exchange strength \cite{Liu_2013}, single-atom impurities \cite{LimaFernandes2018, Hanneken_2016}, and inhomogeneities of the magnetic anisotropy  \cite{Iwasaki2013, doi:10.1063/1.4979316}, can function as pinning sites, slowing down or capturing driven skyrmions in ferromagnetic systems. The same has been shown for current-driven antiferromagnetic skyrmions in a racetrack in presence of a hole \cite{Silva2019} and local variation of the magnetic anisotropy \cite{PhysRevB.100.144439}. Furthermore, the impact of pinning is not limited to current-induced skyrmion dynamics, but has also been observed for the Brownian motion of ferromagnetic skyrmions, only driven by thermal fluctuations \cite{reshuffler, PhysRevB.104.144417,Gruber2022}. While pinning can be detrimental when trying to move them, it can also be advantageous for precisely positioning a skyrmion in a long-term, stable memory device, preventing it from drifting away due to diffusion. 
    
    However, driven ferromagnetic skyrmions display unique characteristics due to a Magnus-type force, typically referred to as \textit{gyrocoupling}, which causes them to move at a certain angle to the driving force. If driven by electric currents, the angle between the skyrmion velocity and the current direction is called skyrmion Hall angle \cite{Nagaosa2013}, in analogy to the conventional Hall effect \cite{Hall1879}. Particle-based simulations have shown that the skyrmion Hall angle is not fixed and depends on the magnitude of the applied driving force in both random \cite{PhysRevLett.114.217202} and periodic \cite{PhysRevB.95.014412, PhysRevB.98.134418} pinning environments, due to the impact of the gyrocoupling being (partly) lifted. 
    In contrast, antiferromagnetic skyrmions do not experience a Hall effect due to being comprised of two interdependent topological objects with opposite topological charges pertaining to each sublattice. When the skyrmion is propelled, the opposing Magnus forces nullify each other, as demonstrated in \cite{PhysRevLett.116.147203,Zhang2016}.  Several studies have investigated the influence of the topological properties of skyrmions on their thermal diffusion, including recent works such as Refs. \cite{PhysRevB.99.224430, PhysRevLett.127.047203, Weissenhofer2022, weiss2020}. The absence of gyroscopic motion in antiferromagnetic skyrmions is also reflected in their thermally-induced motion, which is generally higher than that of ferromagnetic ones \cite{Dohi2022}, which experience diffusion suppression due to their gyrocoupling \cite{PhysRevLett.116.147203,PhysRevB.90.174434}.

    Moreover, suppression of the gyroscopic motion of ferromagnetic skyrmions has also been observed experimentally and in theoretical calculations \cite{doi:10.7566/JPSJ.90.083601,Weienhofer2022Stoch-58039}, where ferromagnetic skyrmions in one-dimensional channels were found to display enhanced diffusion as compared to free diffusion. A similar behavior was reported by Ref. \cite{PhysRevB.104.144417} in micromagnetic simulations of ferromagnetic skyrmion diffusion in granular films. 

    Here we investigate the impact of pinning on the thermally activated motion of ferromagnetic and antiferromagnetic skyrmions. For this purpose, we consider an anti- and ferromagnetically coupled bilayer with defects arising from a local variation of the perpendicular magnetic anisotropy, and discuss the diffusive behavior in the pinning environment created by the defects. Our investigation are conducted through atomistic spin dynamics simulations based on the stochastic Landau–Lifshitz–Gilbert (LLG) equation. 
    We study the impact of both, periodic and random arrangements of defects, and reveal that  ferromagnetic skyrmions experience enhanced diffusion for low Gilbert damping and temperatures and defect strengths near the depinning threshold. Conversely, antiferromagnetic skyrmions generally shows a reduction of thermal diffusion due to pinning. Their behavior is successfully compared to that of a classical Brownian particle in a periodic pinning array based on an equation proposed by Lifson-Jackson \cite{doi:10.1063/1.1732899, RevModPhys.62.251}. The commonalities and differences between the ferromagnetic and antiferromagnetic skyrmion diffusion in the pinning environment are linked to their different topological charges, as can be explained within a rigid-body approach using Thiele's formalism \cite{PhysRevLett.30.230, PhysRevLett.110.127208}.

\section{Methods}

    \label{Methods}

    The system being modeled is a bilayer in which the interactions between atomistic magnetic moments are described via an extended Heisenberg  Hamiltonian \cite{Nowak2007Class-4665}
    \begin{equation}
        \mathcal{H} =  \frac{1}{2}\sum_{i \neq j}  \boldsymbol{S}_i^\text{T}  \mathcal{J}_{ij} \boldsymbol{S}_j    -\sum_{i} d_i{S}_{i,z}^2 \,.
        \label{hamiltonian}
    \end{equation}
    Here,  $i$ and $j$ are indices for nearest neighbor sites and $\boldsymbol{S}_i $ denotes  a unit vector describing a localized magnetic moment. The diagonal elements of the exchange coupling tensor $\mathcal{J}_{ij}$ includes Heisenberg exchange via $J_{ij}=\frac{1}{3}\text{Tr}\mathcal{J}_{ij}$, while the antisymmetric part models the Dzyaloshinsky-Moriya interaction (DMI) $\boldsymbol{D}_{ij} \cdot \left(\boldsymbol{S}_i\times \boldsymbol{S}_j\right)=\frac{1}{2} \boldsymbol{S}_i\left(\mathcal{J}_{ij}-\mathcal{J}_{ij}^\text{T} \right)\boldsymbol{S}_j$. The DMI vectors $\boldsymbol{D}_{ij}$ are situated in the plane and undergo a clockwise rotation around their nearest neighbor sites while also being oriented perpendicular to the connection vector of neighboring lattice sites. 
    The inter-layer exchange parameters are fixed at $J_{ij}=100\,\si{\milli\electronvolt}$, $|\boldsymbol{D}_{ij}|=30\,\si{\milli\electronvolt}$. The intra-layer exchange contributes with $\pm 100\,\si{\milli\electronvolt}$  for ferromagnetic ($+$) and antiferromagnetic ($-$) coupling. Note that the intra-layer DMI is set to zero. The last term introduces uniaxial anisotropy, which is oriented perpendicular to the bilayer. 
    
    The interplay of these interactions leads to the formation of metastable Néel-type skyrmions, which possess a topological charge of $Q=1$ for ferromagnetic coupling and $Q=0$ for antiferromagnetic coupling, respectively. 
    Choosing high values for the interactions is beneficial as it enhances the thermal resistance of skyrmions, making them less susceptible to thermal fluctuations. The ability to perform simulations at elevated temperatures, as a result of increased thermal robustness, makes it more computationally efficient to simulate the diffusion of skyrmions over lengths of multiple lattice constants. 

    To investigate the impact of defects on the behavior of thermally driven skyrmions, a non-homogeneous energy landscape is created by introducing a spatial variation in the magnitude of the anisotropy energy $d_i$. This allows for modeling of atomic defects in the magnetic material, which affects both anti-ferromagnetic and ferromagnetic skyrmions equally. Going forward, $d_{\text{Defect}}$ and $d_0=15\,\si{\milli\electronvolt}$ refer to the  anisotropy energy at defect and non-defect sites respectively.

    The dynamics of the magnetic moments is calculated using the stochastic Landau-Lifshitz-Gilbert (sLLG) equation \cite{LANDAU199251, 1353448}
    \begin{equation}
        \frac{\partial\boldsymbol{S}_i}{\partial t} = - \frac{\gamma}{(1+\alpha^2)\mu_s} \boldsymbol{S}_i\times \left( \boldsymbol{H}_i  +  \alpha \, \boldsymbol{S}_i \times \boldsymbol{H}_i \right) \,,
        \label{sLLG}
    \end{equation}
    with $\alpha$ being the Gilbert damping parameter, $\mu_s$ the atomic magnetic moment, and $\gamma$ representing the gyromagnetic ratio. The local effective field $\boldsymbol{H}_i = -\frac{\partial \mathcal{H} }{\partial\boldsymbol{S}_i} + \boldsymbol{\zeta}_i$ incorporates the contribution from the Hamiltonian as well as a stochastic field $\boldsymbol{\zeta}_i$ accounting for thermal fluctuations. $\boldsymbol{\zeta}_i$ has zero mean  and its autocorrelation is given by 
    \begin{equation*}
        \langle{\zeta}^\mu_i(t) {\zeta}^\nu_j(t')\rangle = 2\alpha\mu_s k_B T \delta_{ij}\delta_{\mu\nu}\delta(t-t')/{\gamma}\, ,
    \end{equation*}
    where $k_B$ is the Boltzmann constant, $T$ is temperature and $\mu,\nu$ denote Cartesian coordinates \cite{PhysRev.130.1677}.

    The numerical integration of the sLLG is accomplished via an GPU-accelerated implementation of Heun's method with a fixed time step of $\Delta t=  \SI{0.1}{\femto\second}$. \cite{Nowak2007Class-4665}. The simulation comprises $64\times 64\times 2$ magnetic moments with periodic boundary conditions along the $x$ and $y$ axis. Following Ref. \cite{PhysRevB.90.174434, PhysRevLett.127.047203, weiss2020, Weissenhofer2022, PhysRevB.107.064423}, the trajectories of skyrmions are obtained by monitoring the out of plane component of the magnetization. 

    Mesoscopically, the motion of localized magnetic textures can be described in terms of a rigid-body approach. The effective equation of motion for ferromagnetic skyrmions, known as Thiele equation \cite{PhysRevLett.30.230}, can be derived from the LLG equation reading 
    \begin{align}
        \boldsymbol{G}\times{\boldsymbol{V}} + \alpha \mathfrak{D} {\boldsymbol{V}} = \boldsymbol{F}\,.
        \label{FM_thiele}
    \end{align}
    Here, $\boldsymbol{V}$ is the velocity of the skyrmion, $\boldsymbol{F}$ represents the force exerted on it, $\boldsymbol{G}=-4\pi Q\mu_s /\gamma a^2 \boldsymbol{e}_\perp$ is the gyrocoupling vector perpendicular to the plane with lattice constant $a$, and $\alpha\mathfrak{D}$ describes dynamic friction. The first term in Eq. (\ref{FM_thiele}) leads to a motion perpendicular to the direction of force, linking the non-trivial topology of the skyrmion to its motion. The second term models dissipation of energy to the heat bath due to its proportionality to the damping parameter. The friction coefficient depends on the specifics of the spin configuration and is calculated via $\mathfrak{D} = \mu_s / (2\gamma a^2)\int (\partial_x \boldsymbol{S}\cdot\partial_x\boldsymbol{S} + \partial_y \boldsymbol{S}\cdot \partial_y\boldsymbol{S})\ \mathrm{d}^2r$ for skyrmions with rotational symmetry \cite{PhysRevB.99.224430, weiss2020}. 

    In order to account for the effect of thermal fluctuations, the force is supplemented with a stochastic force $\boldsymbol{F}^\text{th}$, which has zero mean and an autocorrelation function given by $\langle F_\mu^\text{th}(t)  F_\nu^\text{th}(t') \rangle = 2k_BT\alpha\mathfrak{D}\delta_{\mu\nu}\delta(t-t')$ making Eq. (\ref{FM_thiele}) a Langevin-type equation of motion \cite{Troncoso2014,PhysRevB.97.214426}. The mean-squared-displacement $\langle \left[\boldsymbol{R}(t)-\boldsymbol{R}(0) \right]^2 \rangle = 4Dt$, calculated using an ensemble average over multiple trajectories, allows to determine the free diffusion coefficient of ferromagnetic skyrmions in the absence of external forces \cite{Troncoso2014,PhysRevB.90.174434,PhysRevB.97.214426}:
    \begin{equation}
        D_0^\text{FM}= k_BT \frac{\alpha\mathfrak{D}}{(\alpha\mathfrak{D})^2+G^2}\,.
        \label{D0FM}
    \end{equation}
   Due to the presence of the gyrocoupling term, an unusual relationship between friction and diffusion coefficient is observed. Normally, higher friction leads to decreasing diffusion coefficients, but here, an increasing friction can lead to enhanced diffusion. Throughout this study, the free diffusion coefficient in absence of any potential  is indicated by a zero in the index. 

    In the same way, an equation of motion for antiferromagnetic spin structures can be formulated  reading \cite{PhysRevLett.110.127208}
    \begin{equation}
        M\dot{\boldsymbol{V}} + \alpha \mathfrak{D} {\boldsymbol{V}} = \boldsymbol{F}\,.
        \label{AFM_thiele}
    \end{equation}
    The crucial difference to the ferromagnetic Thiele equation is the absence of the gyrocoupling term. Additionally, the antiferromagnetic Thiele equation includes a mass term with the skyrmion mass $M$, which gives the skyrmion a momentum. This momentum results from the fact that the antiferromagnetic order parameter, the N{\'e}el vector, also experiences inertia \cite{Gomonay2014}. It has been shown \cite{PhysRevB.92.020402,PhysRevLett.116.147203} that antiferromagnetic skyrmions exhibit behavior analogous to a classical massive particle in a viscous medium, as demonstrated by the diffusion coefficient: 
    \begin{equation}
        D_0^\text{AFM}=\frac{k_B T}{\alpha\mathfrak{D}}\,.
        \label{D0AFM}
    \end{equation}
  One can see that the antiferromagnetic diffusivity is generally higher compared to the ferromagnetic skyrmion, since the gyrocoupling in Eq.  (\ref{D0FM}) suppresses the diffusion coefficient. 

In a recent work \cite{PhysRevLett.127.047203} it was demonstrated that the coupling of magnetic textures to thermally excited magnons gives rise to an additional contribution to the damping and the stochastic force. This contribution can be incorporated in the effective equations \eqref{FM_thiele} and \eqref{AFM_thiele} and, subsequently, in Eqs.~\eqref{D0FM} and \eqref{D0AFM}. To achieve this, it is necessary to add a term linear in temperature to the $\alpha\mathfrak{D}$ term. The impact of this magnon-induced friction is most pronounced at high temperatures and low values of the Gilbert damping parameter. Here, however, we neglect this term in the analytical calculations, since for the parameters considered we expect only a minor contribution to the skyrmion dynamics.

The effective diffusion coefficient $D_\text{eff}$ in an arbitrary non-uniform potential cannot be expressed analytically. However, it can be estimated using the Lifson-Jackson equation for the case of a periodic energy landscape, which is relevant for our periodic defect arrangement \cite{doi:10.1063/1.1732899, RevModPhys.62.251}. The Lifson-Jackson equation yields the effective diffusion coefficient of a classical particle in a one-dimensional spatially-dependent periodic potential $U(x)$, whose dynamics is governed by the overdamped Langevin equation, and reads \begin{equation}
D_{\text{eff}} = \frac{D_0}{\langle e^{U/k_BT} \rangle \langle e^{-U/k_BT} \rangle}.
\label{LJeq}
\end{equation} Here, $D_0$ denotes the free diffusion coefficient in
absence of any potential, and $\langle\dots\rangle = \int \dots \text{d}x$ represents the average over one period.

\section{Results}

    \label{Results}

    Having established the theoretical foundations, we now turn to the results of our analysis. First, we investigate the diffusion of single skyrmions in the presence of atomic defects placed periodically on a grid with a spacing of four lattice constants. To put that into perspective, the diameter of a skyrmion is  roughly 13 lattice constants.
    
\begin{figure}
        \centering
        \begin{subfigure}
            \centering
            \includegraphics[width=0.37\linewidth]{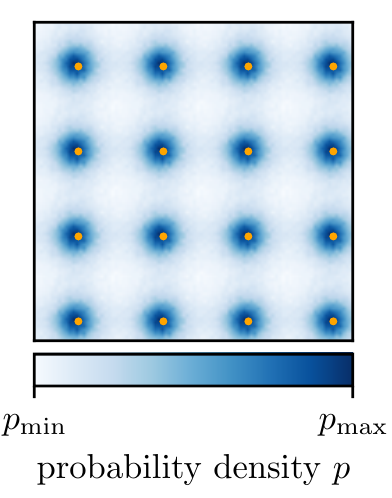}
        \end{subfigure}
        \hspace{-10pt}
        \begin{subfigure}
            \centering
            \includegraphics[width=0.63\linewidth]{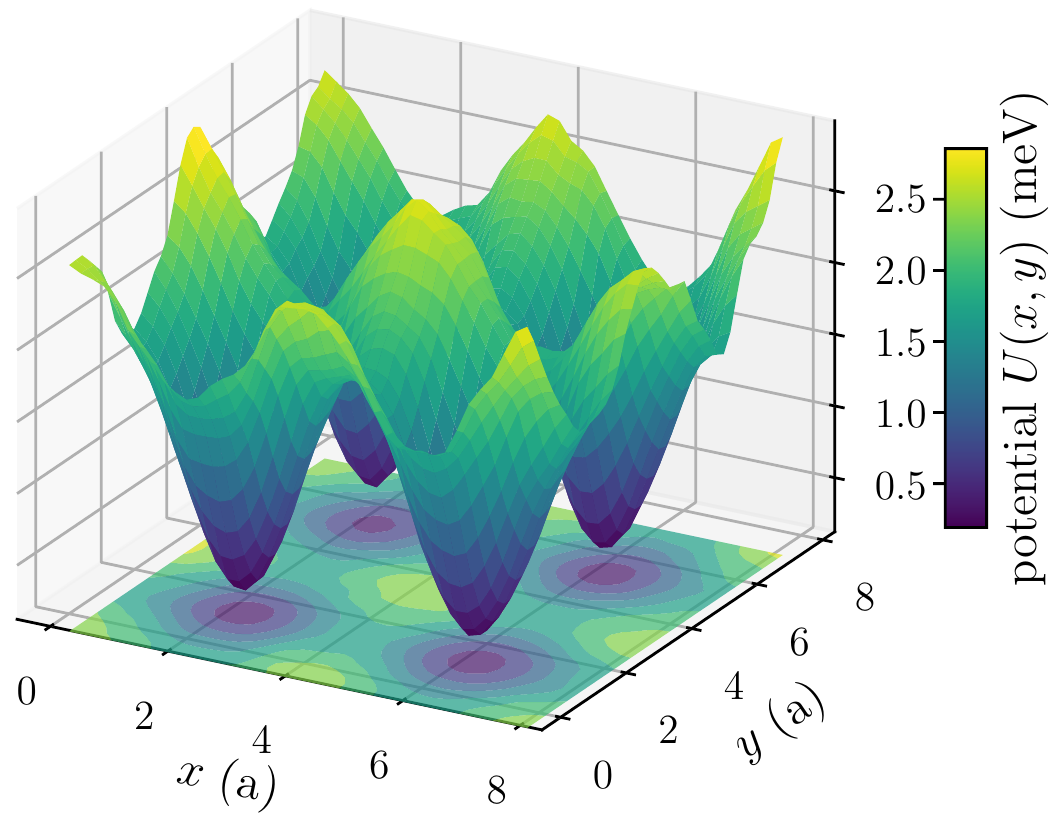}
        \end{subfigure}
        \caption{Probability density representing the likelihood of finding a skyrmion at a given point (left) and a small section of the corresponding energy landscape (right) of the periodic defect arrangement. The orange dots indicate defect locations and have  a distance of 4 lattice constants. Simulations were performed at $\alpha=0.1$ and $k_BT=1\,\si{\milli\electronvolt}$ and a defect anisotropy reduced by a factor of  $d_{\text{Defect}}/d_0=0.9$ or $d_0 - d_\text{Defect} = 1.5\,\si{\milli\electronvolt}$.}
        \label{fig:energy_landscape}
    \end{figure}

    An analysis of the positional occupation statistics of diffusing skyrmions can provide insights into the defect-induced potential \cite{Gruber2022}. This is because the population probability histogram $p(x, y)\mathrm{d}x\mathrm{d}y$, which represents the probability of finding a diffusing skyrmion in a small area around the location $(x, y)$, is related to the potential via the Boltzmann distribution, $U(x,y)\propto-k_BT \operatorname{ln}p(x,y)$ \cite{Evans1979}.
    In Fig.~\ref{fig:energy_landscape}, a small section of the computed probability density is displayed on the left. By comparing the positions of the defects, represented by orange dots, with the probability density, it can be seen that a skyrmion tends to congregate around these defects, which demonstrates their pinning effect. The energy landscape (with $\mathrm{min}(U)=0$) of a section with 4 defects is displayed on the right, with a peak height of $\SI{2.5}{\milli\electronvolt}$ right between next-nearest neighbor defects. Adjacent pinning sites are most easily transitioned along the $x$ or $y$ directions, with an energy barrier of $\SI{1.5}{\milli\electronvolt}$.
    Note that for the parameters chosen here, the anisotropy energy difference between a defect and a normal site is also $\Delta d=d_0- d_\text{Defect}=\SI{1.5}{\milli\electronvolt}$ or $d_\text{Defect}/d_0 = 0.9$.  If the thermal energy is not significantly higher than this energy barrier, effectively only four escape paths remain for the diffusing skyrmion, due to the exponential behaviour of the depinning process.

\begin{figure}
        \centering
        \includegraphics[width=1\linewidth]{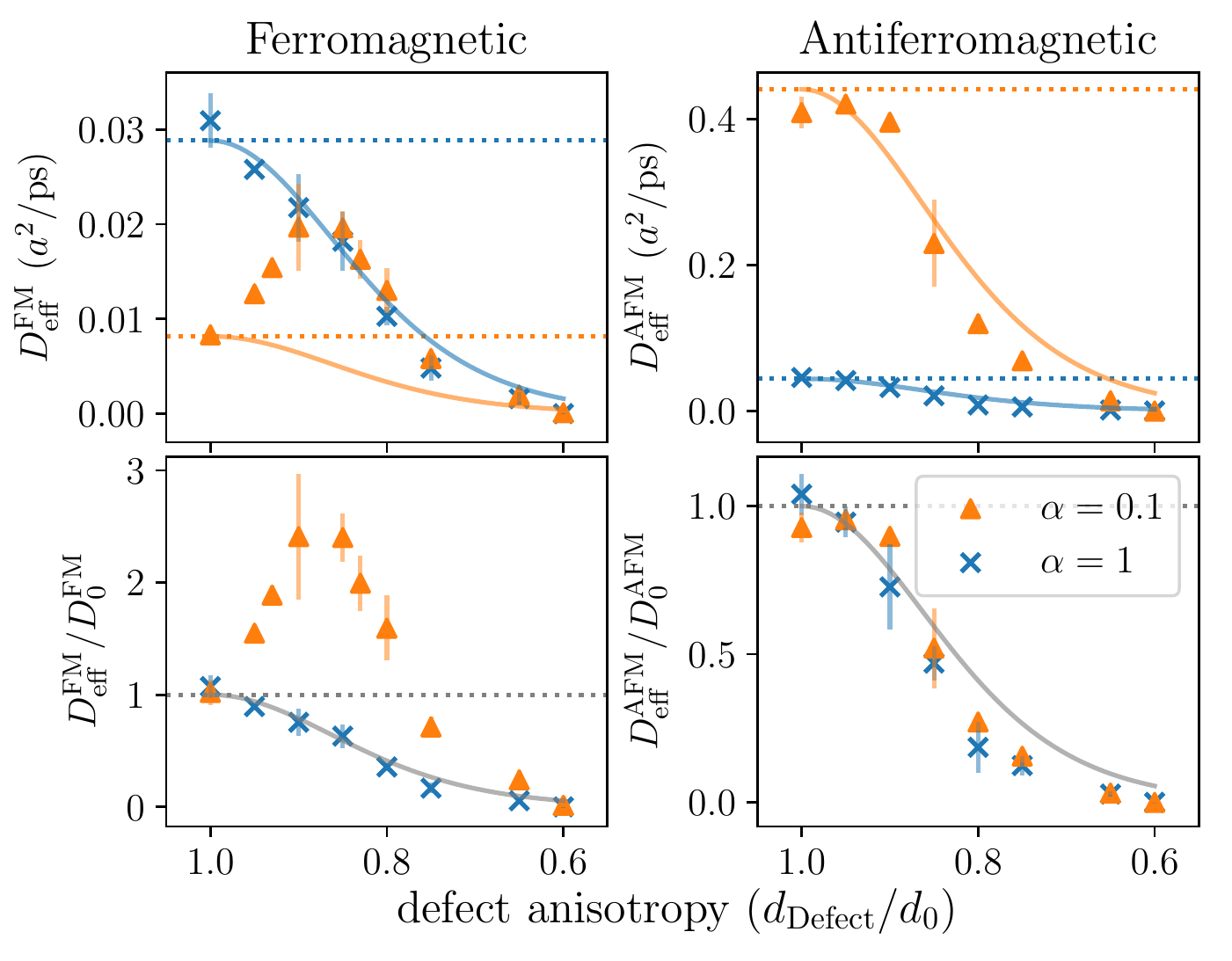}
        \caption{Anti- and ferromagnetic diffusion coefficients versus defect strength for various $\alpha$. Defect strength is expressed by the ratio between the defects anisotropy energy $d_\text{Defect}$ and the global anisotropy energy $d_0$. Symbols represent simulation results at $k_BT = 1\,\si{\milli\electronvolt}$ in a periodic defect configuration $\alpha$ as labeled. The diffusion coefficients on the bottom graphs are normalized by the expectation of free diffusion, see Eqs. (\ref{D0AFM}) and (\ref{D0FM}). Data is compared to free diffusion (dotted lines) and the Lifson-Jackson equation (solid lines). }
        \label{fig:defect_strength_vs_D}
    \end{figure}

    The presence of the defect potential is also manifested in the diffusion coefficient of the skyrmions, which determines their thermal mobility. This is demonstrated in Fig.~\ref{fig:defect_strength_vs_D}, where the dependence of the diffusion coefficient on the defect strength is depicted. The defect strength is controlled by altering the amplitude of the anisotropy energy $d_\text{Defect}$ at the defect's locations, and the horizontal axis is expressed in terms of the ratio between the defect's anisotropy energy and the uniform anisotropy energy $d_0$ of the surrounding atoms. This means that a lower ratio $d_\text{Defect}/d_0$ corresponds to a higher defect strength.  It is also important to note that the diffusion coefficients in the lower plot have been normalized by their respective free diffusion coefficients according to equations (\ref{D0AFM}) and (\ref{D0FM}). Normalizing the data in this way allows for a direct comparison of the effective diffusion coefficients in the presence of defects to those in the absence of defects.  
    
    The diffusion coefficients in absence of defects ($d_\text{Defect}/d_0 = 1$) are consistent with the analytical predictions for free diffusion as given in equations (\ref{D0FM}) and (\ref{D0AFM}) and represented by the dotted lines for reference. Apart from that there is a significant difference in the diffusion coefficients of ferromagnetic (left) and antiferromagnetic (right) skyrmions. The diffusivity of antiferromagnetic skyrmions decreases continuously as the defect strength increases. Besides, the ratio of diffusivity to free diffusion seen in the lower right plot is independent of $\alpha$. The similarity to a classic Brownian particle exhibiting decreased mobility due to pinning is evidenced by the solid line representing the expectation of the Lifson-Jackson equation \ref{LJeq}. The latter is evaluated from the one-dimensional escape paths seen in the potential in Fig. \ref{fig:energy_landscape} and using the proportionality between the defect anisotropy and energy landscape. 
    
    The behavior of ferromagnetic skyrmions in the left part of Fig.~\ref{fig:defect_strength_vs_D} deviates from this trend, displaying peculiar characteristics. Unlike the antiferromagnetic skyrmion, the degree to which the ferromagnetic diffusivity is affected appears to depend on the damping.  When the damping is low ($\alpha=0.1$), the ferromagnetic diffusion coefficient of the skyrmion initially increases with increasing defect strength before eventually converging to zero. The observed rise in thermal mobility above the expectation for free diffusion suggests that the typical diffusion suppression \cite{PhysRevB.90.174434} caused by gyrocoupling in ferromagnetic skyrmions is partially counteracted in this case. For high damping ($\alpha=1$), the increase in mobility does not occur. Instead, the diffusivity of the ferromagnetic skyrmion also continually decreases as the defect strength increases, and its behavior is well described by the Lifson-Jackson equation.

\begin{figure}
        \centering
        \includegraphics[width=1\linewidth]{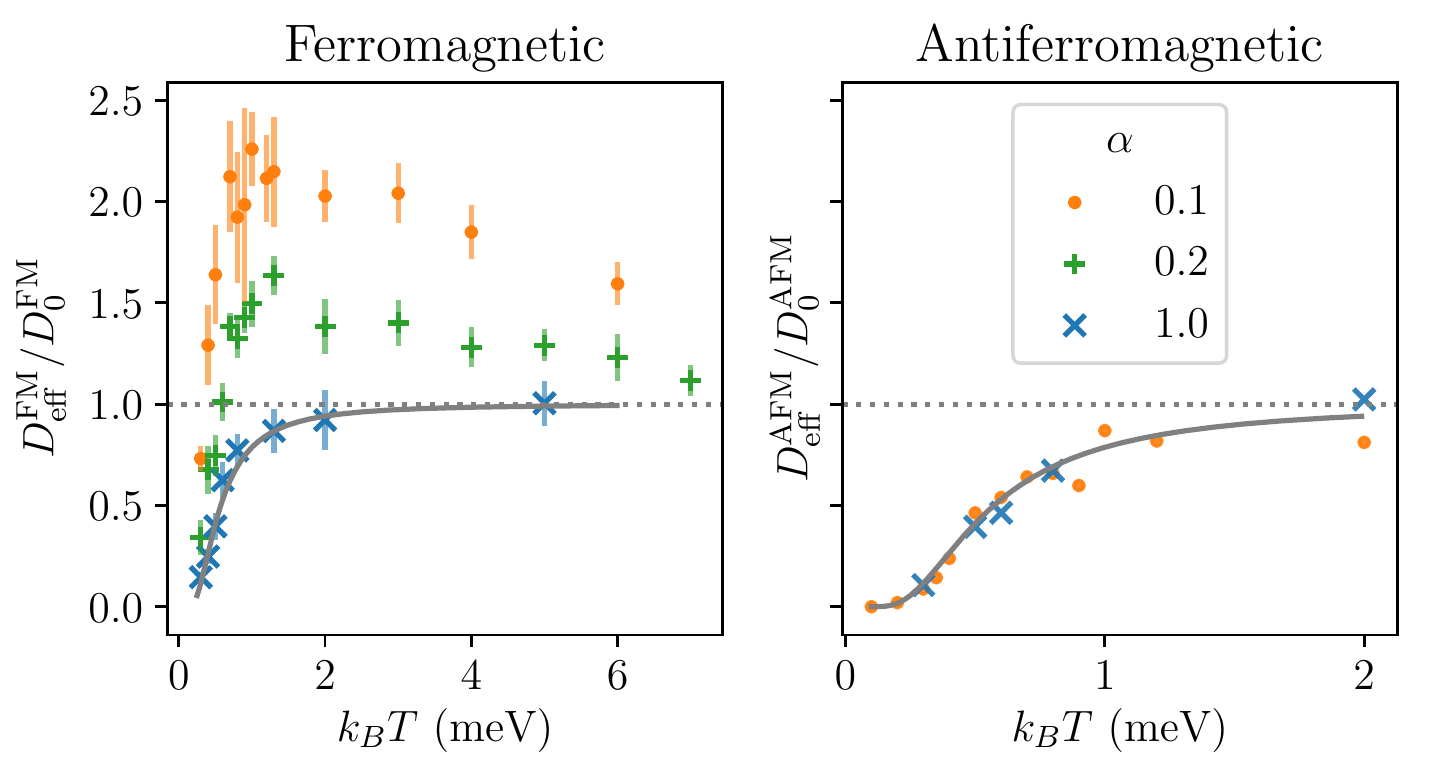}
        \caption{Diffusion coefficients of anti- and ferromagnetic skyrmions in the periodic defect configuration versus thermal energy for various $\alpha$ as labeled. The diffusion coefficients are normalized by their respective  free diffusion coefficients according to Eqs. (\ref{D0AFM}) and (\ref{D0FM}). The prediction of the Lifson-Jackson equation (\ref{LJeq}) is represented by the solid line. Simulations performed with $d_\text{Defect}/d_0 = 0.9$. }
        \label{fig:kT_vs_D}
    \end{figure}

    Instead of changing the underlying energy landscape by altering defect strength, we hold defect strength constant but change thermal energy, which also determines the influence of the defects on the skyrmion. The dependence of diffusivity on temperature can be seen in Fig.~\ref{fig:kT_vs_D} for anti- and ferromagnetic skyrmions. Again, the effective diffusion coefficients in presence of defects are shown in terms of the relative deviation from free diffusion. One can see that the antiferromagnetic skyrmion displays reduced diffusivity across the board. 
    The agreement between the prediction of the Lifson-Jackson equation \ref{LJeq} and the simulation results suggests that -- as long as the Gilbert damping parameter is not too small --  antiferromagnetic skyrmions behave like classical particles in a viscous medium, the dynamics of which are governed by the overdamped Langevin equation \cite{RevModPhys.62.251}. 
    
    Ferromagnetic skyrmions follow this trend only at high damping, $\alpha = 1$. For lower values of $\alpha$, its diffusion coefficient exceeds the expected value for free diffusion, rather than falling below it. One can see that this is more pronounced as $\alpha$ decreases. Note that the increase in diffusivity is at its largest around $k_\mathrm{B}T =\SI{1.5}{\milli\electronvolt}$, which coincides with the energy barrier between two neighbouring defects (see Fig. \ref{fig:energy_landscape}). This behaviour is similar to what was recently observed for domain walls in ferromagnets, where the maximum and the subsequent drop in the diffusion coefficient with rising temperature was interpreted as a \textit{Walker breakdown of Brownian domain wall dynamics} \cite{PhysRevB.106.104428}

\begin{figure}
    \centering
    \includegraphics[width=0.7\linewidth]{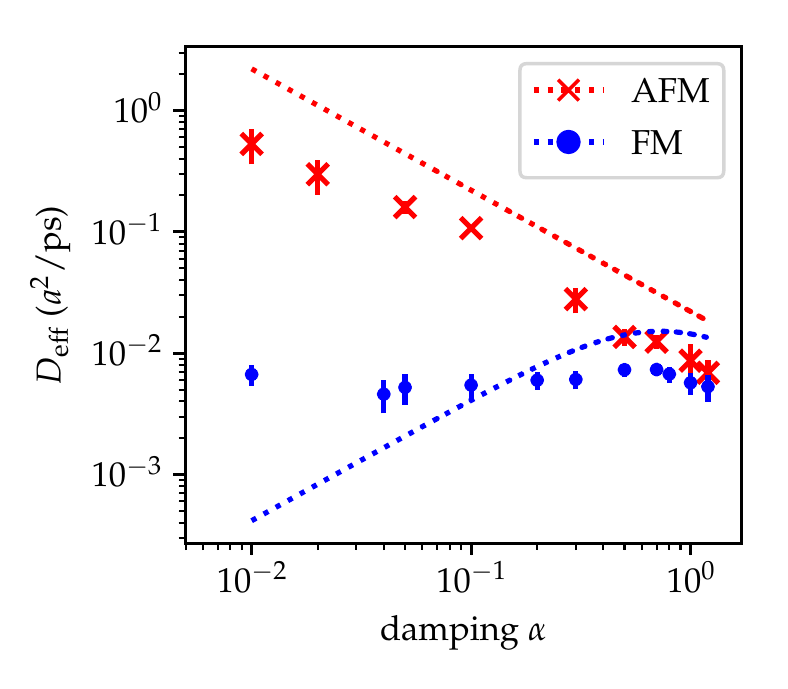}
    \caption{Anti- and ferromagnetic diffusion coefficients versus $\alpha$. Diffusion coefficients obtained from simulations performed at $k_BT=1\,\si{\milli\electronvolt}$ and $d_\text{Defect}/d_0 = 0.9$. Dotted lines indicate the prediction of free diffusion according to Eqs. [\ref{D0AFM}] and [\ref{D0FM}].}
    \label{fig:alpha_vs_D}
\end{figure}

    The dependence on damping is more closely investigated in Fig. \ref{fig:alpha_vs_D}, where one can see diffusion coefficients of anti- and ferromagnetic skyrmions for a range of different damping parameters in the periodic pinning array. It appears that the ferromagnetic diffusion coefficient does not  change significantly over three orders of magnitude of the damping parameter. The effect of enhanced diffusion increases as $\alpha$ decreases when comparing the effective diffusion coefficients in the periodic defect array to free diffusion (dotted line). However, the ferromagnetic diffusion suppression is not entirely lifted as the diffusivity still falls way below the antiferromagnetic curve.

    An explanation to why the anti- and ferromagnetic skyrmions behave either similar or completely opposite depending on damping is provided by Thiele's equations (\ref{FM_thiele}) and (\ref{AFM_thiele}). As damping increases, the mass and gyrocoupling terms become less significant, as they are not dependent on $\alpha$, unlike the friction term. As a result, the dynamics of both the anti- and ferromagnetic skyrmions are effectively controlled by the same equation of motion in the overdamped limit. However, in the case of low damping, the gyrocoupling term, which is only present in the ferromagnetic Thiele equation, becomes more significant and starts to have a greater impact on the dynamics, leading to a deviation from the antiferromagnetic skyrmion. The transition region at which this deviation occurs can be approximated by examining the ratio between $\mathfrak{D}$ and $ G$, which is roughly $1$ for the skyrmions under consideration, and can be roughly seen in Fig. \ref{fig:alpha_vs_D}. However, our simulations revealed that this difference is less pronounced when skyrmions are subject to pinning. An explanation is given by considering the defect-induced energy landscape in Fig.  \ref{fig:energy_landscape}. Since the jumps between pinning sites preferably occur where the energy gap between them is smallest, skyrmions tend to move in one-dimensional channels.  As a result, the gyroscopic motion of ferromagnetic skyrmions cannot fully develop and its effects are impeded, meaning that the suppression of the diffusion is partly lifted. The resulting increase in diffusivity outweighs the usually hindering effect imposed by the defects as seen in Fig. \ref{fig:defect_strength_vs_D} for certain defect strengths. The diffusion in one-dimensional channels has been studied experimentally in Ref. \cite{doi:10.7566/JPSJ.90.083601}, where it was observed that ferromagnetic skyrmions exhibit increased diffusion under this constraint.

    This effect is similar to what is observed from a skyrmion under an applied active drive. In the absence of any obstructions, the Hall angle of a driven skyrmion remains unchanged, regardless of the strength of the driving force \cite{nnano.2013.243, PhysRevLett.107.136804}. However, when the skyrmion encounters single, periodic, or random defects, its Hall angle becomes more complex. As reported in Refs. \cite{PhysRevB.91.054410,PhysRevB.91.104426, PhysRevLett.114.217202} using both, continuum and particle-based approaches, the Hall angle of the skyrmion is at its lowest when the driving force reaches the depinning threshold. Beyond this point, the angle gradually increases with each increase in driving force, before reaching a saturation point at the free pinning angle. This behavior has been observed in previous experimental studies (Ref. \cite{Jiang2017,Litzius2017,Litzius2020}). The Magnus force causes a skyrmion to be redirected as it moves through a pinning site, resulting in a change in direction towards the driving force. As the applied drive increases, the skyrmion moves more quickly reducing the extent of the change in direction. This can result in skyrmion motion that exceeds the velocity attainable from the applied forces alone \cite{PhysRevB.91.104426}.

    The change in behavior of thermally-driven ferromagnetic skyrmions in the presence of pinning is in line with previous observations \cite{doi:10.7566/JPSJ.90.083601,PhysRevB.104.144417,Weienhofer2022Stoch-58039}. Our results, displayed in Figures \ref{fig:defect_strength_vs_D} and \ref{fig:kT_vs_D}, show a noticeable increase in diffusivity, which suggests a decrease in the Magnus force. The greatest suppression of the Magnus force is seen when the thermal energy matches the depinning threshold at $1.5\si{\milli\electronvolt}$, as shown by comparing the temperature dependence of the diffusion coefficient in Fig. \ref{fig:kT_vs_D} to the energy landscape in Fig. \ref{fig:energy_landscape}.

    In contrast to ferromagnetic skyrmions, antiferromagnetic skyrmions do not exhibit the skyrmion Hall effect \cite{PhysRevLett.116.147203}. As demonstrated in Ref. \cite{PhysRevB.100.144439}, current driven antiferromagnetic skyrmions interacting with a vacancy tend to slow down, and this is also observed for the thermally driven antiferromagnetic skyrmions investigated here.

    Our findings discussed so far demonstrate that ferromagnetic skyrmions exhibit enhanced diffusion in a periodic defect environment. However, real magnetic thin films are unlikely to have a perfectly ordered arrangement of atomic defects. To account for this, we are abandoning the assumption of periodicity and instead looking at defects that are distributed randomly throughout the bilayer. We have chosen the probability that a lattice site is a defect in a way that maintains the same overall defect density as before in the periodic defect arrangement. To calculate the diffusion coefficient, we calculate the mean squared displacement as an ensemble average of different skyrmion trajectories, as previously mentioned. However, in this case, we utilize different, randomly generated defect configurations for each trajectory.
    
\begin{figure}[]
        \centering
        \begin{subfigure}
            \centering
            \includegraphics[width=0.52\linewidth]{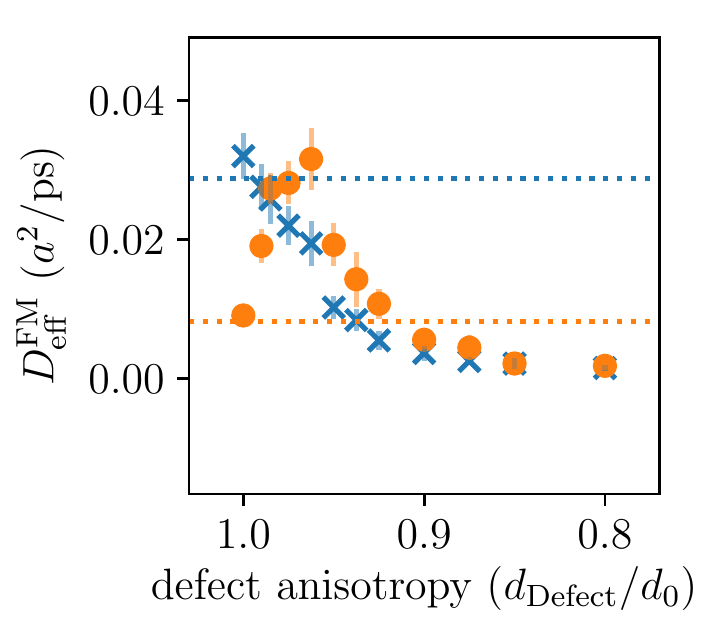}
        \end{subfigure}
        \hspace{-14pt}
        \begin{subfigure}
            \centering
            \includegraphics[width=0.50\linewidth]{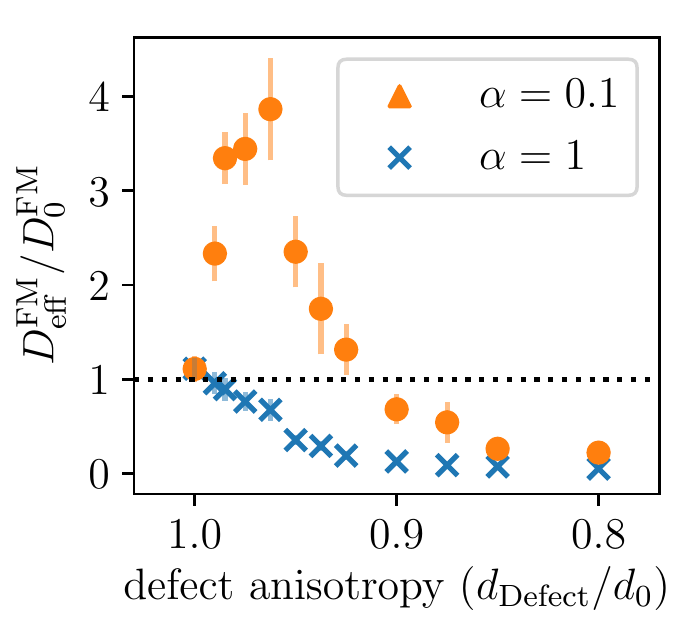}
        \end{subfigure}
        \caption{Diffusion coefficients versus defect strength of ferromagnetic skyrmions  in presence of randomly distributed defects. Symbols represent simulation results at $k_BT = 1\,\si{\milli\electronvolt}$ and $\alpha$ as labeled. The diffusion coefficients on the right graph are normalized by the expectation of free diffusion, see Eq. (\ref{D0FM}).}
        \label{fig:diff_c_rand}
    \end{figure}

    Fig.~\ref{fig:diff_c_rand} depicts the dependence of the ferromagnetic diffusion coefficient  on defect strength in a random pinning environment. All trends coincide with the previous observation of ferromagnetic skyrmions in a periodic defect array seen in Fig.~\ref{fig:defect_strength_vs_D}. Therefore, the phenomenon of enhanced diffusivity is not limited to periodic defect patterns but is present in general pinning environments.

    \section{Conclusion}

    In this study, we investigated the behavior of ferro- and antiferromagnetic skyrmions under periodic and random pinning conditions. Our results reveal that, in contrast to normal Brownian particles, pinning can enhance the thermal mobility of ferromagnetic skyrmions at low Gilbert damping. This effect depends on the strength of the pinning defects and temperature, with the greatest increase in diffusion occurring when the thermal energy coincides with the energy of transition between pinning sites. For high damping, the ferromagnetic skyrmion exhibits reduced diffusion only. In contrast to this, the Brownian motion of antiferromagnetic skyrmions in a periodic pinning environment is consistent with the Lifson-Jackson model, which predicts reduced diffusion independent of damping. Thiele's equations of motion can explain the similarities and differences in diffusion behavior between anti- and ferromagnetic skyrmions based on their distinct topological properties. The reason why damping plays a crucial role can be traced back to the fact that the gyrocoupling and mass term, which define the distinct characteristics of anti- and ferromagnetic skyrmions respectively, become more prominent with higher damping and less pronounced with lower damping. These findings align with previous observations of ferromagnetic skyrmions driven by current rather than temperature \cite{PhysRevLett.114.217202,PhysRevB.95.014412, PhysRevB.98.134418}, indicating that under the influence of defects, the Hall angle or gyroscopic motion is dependent on the active drive and the Magnus force can be nearly entirely suppressed at the depinning threshold.


\begin{acknowledgments}
We acknowledge funding by the Deutsche Forschungsgemeinschaft (DFG, german research foundation) via SFB 1432 and project number 403502522.

\end{acknowledgments}

\bibliographystyle{apsrev4-1}

\bibliography{bibfile}

\end{document}